# 3LSAA: A Secure And Privacy-preserving Zero-knowledge-based Data-sharing Approach Under An Untrusted Environment

Wei-Yi Kuo    Ren-Song Tsay

Logos Advanced System Lab,

Dept. of Computer Science, National Tsing-Hua University, Taiwan

**ABSTRACT**

As data collection and analysis become critical functions for many cloud applications, proper data sharing with approved parties is required. However, the traditional data sharing scheme through centralized data escrow servers may sacrifice owners' privacy and is weak in security. Mainly, the servers physically own all data while the original data owners have only virtual ownership and lose actual access control. Therefore, we propose a 3-layer SSE-ABE-AES (3LSAA) cryptography-based privacy-protected data-sharing protocol based on the assumption that servers are honest-but-curious. The 3LSAA protocol realizes automatic access control management and convenient file search even if the server is not trustable. Besides achieving data self-sovereignty, our approach also improves system usability, eliminates the defects in the traditional SSE and ABE approaches, and provides a local AES key recovery method for user's availability.

## I. Introduction

In this paper, we propose a 3-layer cryptography-based fully privacy-protected data-sharing system protocol that allows automatic access control management with convenient file search feature and great usability improvement. Our system protocol is aiming at general users for secure data sharing even under an untrusted environment.

People are producing enormous data daily, in both the virtual and real world, and these data are of interest to many parties for various purposes. For example, patients can be *data owners* and produce personal health records (PHRs) from clinical services. At the same time, doctors or research institutions can be *data users* who may access and analyze these PHRs for precise diagnosis and treatment [10]. Generally, proper data sharing with approved parties can be beneficial. However, data privacy protection from unauthorized parties without sacrificing usability is one most critical issues to resolve for building a secure data sharing ecosystem. In the past decades, significant progress has been made on this issue, but the solutions are far from satisfactory.

As for privacy protection, an intuitive secure solution is to have every data owner manage his own data storage and access control. However, in addition to the stringent accessibility and reliability requirements, and the potentially unbearable cost of data storage, the computation load of authentication and authorization can easily overwhelm regular data owners.

In contrast to the personal data management solution above, the centralized cloud server solution is far more economical than privacy-sacrificing. Essentially, a data owner may store her data in a selected cloud server, which plays the role as a data escrow agent. The server then helps to manage authentication and authorization for the original data owners, and provides highly accessible and reliable data storage to facilitate many-to-many transactions [13]. A typical usage model is that a user with an owner-authorized access token can access the data stored in the escrow agent. Compared to personal data management, the cloud server solution is indeed more practical for data transaction management. However, one major issue is that in this scheme, the server physically *owns* the stored data, and the original owners become *virtual owners* in reality and lose the total access control. If the server is hacked or is dishonest, unauthorized access may occur and infringe the privacy of data owners [12]. Additionally, since the server performs all the physical data accesses, it may track and analyze the access behaviors to reveal data owners and users' personally identifiable information (PII) [9].

Previously, some have introduced a CIA (confidentiality, integrity, and availability) triad model [17] to evaluate the data privacy and security level within a trusted organization. Nonetheless, the cloud-based data transaction services generally operate in untrusted environments across organizations, with widespread owners, users, or even cloud servers. We hence propose a new extended CIA (ECIA) model to cover the dishonest behaviors from possible participants. Here, we adopt a realistic assumption to simplify the discussions. Since protocol violations benefit no one, the rational choice for each legal participant is to be an *honest* protocol executor but still be *curious* about learning extra information or trying to poke and peek. Under this honest-but-curious [12] assumption, we exclude the possibility that the cloud server may deliberately delete data or return the wrong results. However, we simply call those *curious* servers as being potentially *dishonest* for later discussions.

The traditional CIA confidentiality measure concerns only data privacy and security protection from unauthorized access. Nevertheless, with the possibility of *dishonest* servers, the confidentiality measure shall be extended to cover illegal authorizations in addition to merely unauthorized accesses.

For the data owners to protect the data privacy and security in the case of data escrow under dishonest cloud servers, data encryption is an effective way to take back the actual data ownership. However, the server still controls the physical ownership. In practice, an owner may encrypt each piece of her data using a unique Advanced Encryption Standard (AES) key before uploading the encrypted data to the cloud server. By applying data-specific keys, the risk of key leakage is greatly minimized. Most importantly though, the enforced data encryption effectively protects the confidentiality of data content and the owner's privacy. By controlling keys, and having only encrypted data on the server, the data owner regains the true ownership, as physical data accesses are useless unless the decryption keys are available.

However, data encryption introduces a new challenge to the traditional keyword search, which requires plaintext keywords to match data contents. Note that the keyword search is an essential usability tool for a user to locate the target data, as no one has the capability and capacity to remember the contents of all data. One solution is to tag each piece of data with a set of searchable keywords. Nevertheless, the tagged keywords shall be encrypted in some way to achieve extended confidentiality so that no one can extract privacy-related information, such as linking an owner's identification to the blood pressure value matched by the keyword.

Therefore, the search algorithm should be enhanced so that the server cannot identify the actual search keyword but can still locate the matched data. Additionally, the search results shall reveal no linkage between the keywords and the matched data. For this purpose, a secure-searchable encryption (SSE) approach based on the zero-knowledge concept was proposed to facilitate secure search requests on encrypted data. Essentially, the SSE approach applies bilinear map encryption [1][2] on both the search keywords and the data-tagged keywords. Such an owner secret key encrypted search keyword is usually named a search token. Through the unique property of bilinearity, the server can then compute whether the search tokens match any data-tagged tokens and respond to the matched data to the user while knowing nothing about what is being matched.

Although working reasonably well for encrypted searches, the SSE method has a shortfall. Mainly because that the SSE search token is user neutral, any party who holds a search token can repeat the same search. Hence, a clever Attribute-Based Encryption (ABE) method was developed to verify the authentication proof while revealing no identification of the user [3][4][7]. The idea is to delegate to the server an owner-specified fine-grained access control policy encrypted in a bilinear map format, and have Attribute Authorities (AAs), each to qualify a specific fine-grained attribute, or specific role, of the token-user. For instance, a hospital can be an AA to authenticate its employed doctors, and a school may authenticate its enrolled students. Each AA issues to the token-user a bilinear map encrypted attribute credential (AC). Note that by delegating the attribute authentication tasks to AAs, the data owner is alleviated from complicated authentication tasks.

Consequently, an ABSE approach, combining the SSE search and ABE authentication methods, was proposed to qualify the user for token search. In practice ABSE has two possible ways of executing the SSE search and ABE authentication. One is to perform SSE before ABE [6], and the other is in the reversed order [15][20]. Since the computation of one search token matching is more efficient than that of an access policy verification, the SSE-before-ABE approach is preferred. For ABSE implementation, the server applies a bilinear mapping computation to qualify the user's collected attribute credentials against the owner-specified access policy after search token matching. In other words, the server restricts only an ABE-qualified user to receive the SSE-matched data and hence remedies the issue of random reuse of search tokens. Since the computation of bilinear mapping encryption is a zero-knowledge process, the ABSE does not compromise the identity privacy of users and achieves confidentiality.

Although the ABSE method [6][15][20] does alleviate the extended confidentiality concern, it still contains several vulnerabilities and requires enhancements to meet the ECIA condition. First, the ABSE is merely the integration of SSE-ABE or a 2-layer encryption method, which does not specifically address the issue on how to decrypt the returned AES-encrypted data. We regard as the AES encryption on returned data as the third layer encryption, which is integrated to a 3-layer SSE-ABE-AES (3LSAA) system. Intuitively, the user, after receiving data, may contact the data owner for the AES decryption key. However, unlike servers, individuals are not always available. Therefore, we propose an AES local decryption approach that lets the qualified user personally recover the decryption key based on her local proof of authentication after receiving matched data through the ABSE process. Note that we ensure recovery of corresponding AES key based on the same ABSE process that identifies the matched data but using a different set of tokens. Essentially, the third AES layer lets users perform local AES decryption after receiving matched data.

To realize the local decryption, basically, on top of the owner-consent search token and AA authenticated credentials required by ABSE, the owner also issues the user a decryption token. The user then applies the decryption token to the received data from ABSE search and recovers the AES key locally for data decryption. Note that the local key recovery method prevents servers from deliberately reading data contents and preventing unauthorized users from decrypting the received data even if someone somehow delivers it to them.

Besides the contribution of our 3-layer integration with token search and local decryption, we also provide solutions raised by Lewko et al [4] for the collusion issue that occurs in the traditional ABE method [3]. Mainly, someone may combine person A's student attribute credential and person B's employee attribute credential, for example, to cheat the system since the server cannot tell the ownership of the attribute credential. Lewko's proposed method associates each authenticated attribute credential with a user-specific global identity (GID). It enables the server to verify if all attribute credentials are from the same qualified user. However, since users use the same GID for each search, Lewko's method can easily leak out user identity after second requests. Hence, we propose a solution that creates a temporary user attribute credential using a nonce instead of a fixed value. Therefore, the server cannot link the request to user identity with different nonce in each request.

One issue of the SSE search [1] is that its search token locates indiscriminately all data that contain the supplied keyword, while practically the owner may want to confine search range. Therefore, we integrate Cui's method [2] into our SSE layer for a more flexible search that allows the user to search only a partial set of the data, instead of the whole set. After first classifying their data into $n$ sets and using $i \in N = \{1, 2, \ldots, n\}$ for data set index, owners may confine a search token to partial data sets of indices in $S \subset N$. Cui's method requires an index-specific search token for each piece of data. Our improvement over Cui's method is that we require only one search token for searching data in the subset S, and the search computation time is reduced n times for n pieces of data.

Practically, in addition to the extended confidentiality features discussed above, for better usability, a data owner needs to be able to update data properties, which include data-tagged keywords, access policy, and encryption key. But data properties update causes integrity issues. In contrast, the traditional integrity requirement permits no unauthorized data deletion or modification but ensures needed recovery. Now for the cloud server case, the extended integrity should grant specifically only the data owner the right to update the data properties. However, since all data properties are encrypted, the update process needs to be performed under encrypted form. We hence propose that in such a re-encryption update, the data server follows the ABSE authentication method and allows only the authenticated owner for re-encryption update. Note that the re-encryption process is based on zero knowledge proof and the server cannot identify the owner.

In summary, our contribution is threefold. First, we extend the CIA triad model to cover the untrusted cloud server data sharing system and also well address the usability concern. Second, the most unique feature of our approach is the local user AES data decryption method which resolves both the owner availability issue and data security issue. Third, we devise a 3-layer approach that greatly enhances search flexibility, storage efficiency, privacy protection and is effectually anti-collusion.

The rest of this paper is organized as follows. In Section 2, we review some preliminaries used in our construction. Section 3 formalizes the system model into the basic and advanced parts and describes a concrete construction. Section 4 discusses the security proof and complexity of our proposed scheme. Finally, section 5 gives a brief conclusion.

**II. Preliminaries**

In this section, we elaborate the bilinear mapping method which is the main encryption method used in this paper and discuss the security assumption. The effectiveness of our proposed 3LSAA approach is based on the zero-knowledge proof following the bilinearity property of the bilinear mapping method. Additionally, we assume that no existing polynomial-time algorithm can efficiently solve the computational hardness of cryptographic primitives of provable security [19]. In other words, a system is secure if every adversary is of limited computation capability. In the following, we discuss details.

2.1 Bilinear map

Let G and GT denote two cyclic multiplicative groups with prime order $q$, $Z_q = \{0, 1, 2, \cdots, q-1\}$ and $g$ be a generator of G. Define a mapping function on G and GT as e: G × G → GT. If the mapping function $e$ satisfies the following properties, then $e$ is a bilinear map function.

(1) Bilinearity property
- For all $u, v \in Z_q$ and $x, y \in G$, then $e(x^u, y^v) = e(x, y)^{uv}$
- For all $x_1, x_2, y \in G$, then $e(x_1 x_2, y) = e(x_1, y)e(x_2, y)$
- For all $x, y_1, y_2 \in G$, then $e(x, y_1 y_2) = e(x, y_1)e(x, y_2)$

(2) Computability property

For all x, y ∈ G, a polynomial-time algorithm calculates $e(x, y) \in$ GT efficiently.

(3) Non-degenerate property

$e(g, g) \neq 1$, where 1 is the identity of G.

2.2 Zero-knowledge proof with bilinearity

The zero-knowledge proof, a concept from cryptography, is a method that one party can prove to another that a mathematical statement is true, without revealing anything other than the veracity of the statement. In our system, we utilize the bilinearity property to implement the zero-knowledge proof. In Fig. 1, we take the encrypted data search as an example [1] to illustrate the process.

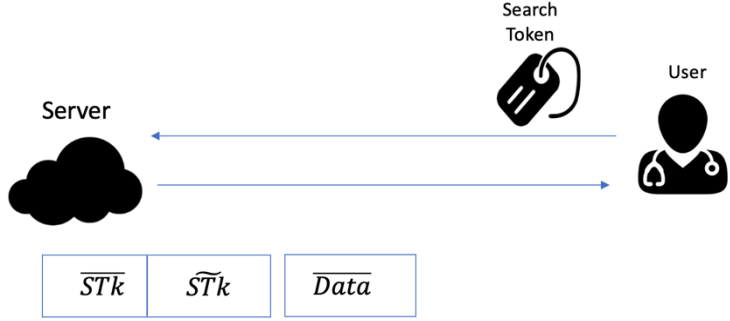

Fig. 1. The application of zero-knowledge proof in the SSE encrypted database search.

For a user SSE zero-knowledge encrypted search to locate the data, he shall first obtain an encrypted keyword search token from owner, $\overline{STk} = H(w)^{sk}$, where $sk$ is owner's secret key, $w$ is a keyword, and $H(.)$ is a hash function. Then send $\overline{STk}$ to the server. For the rest of this paper, we will use the overlined notation to indicate an encrypted element. Note that for each piece of data the owner shall set up the environment by first picking a nonce $r$ and installing in the server a search token transferor, $\widetilde{STk} = g^{\frac{r}{sk}}$, and an encrypted keyword, $\overline{kw} = e(H(w), g)^r$. Here, we use the tilde notation to denote a cryptography element transferor. For example, $\widetilde{STk}$ is for $\overline{STk}$ transfer computation which is to be explained later Finally, the owner chooses a data-unique AES key to encrypt the original data into a ciphertext, i.e., $\overline{Data}$, which is then sent to the server along with $\widetilde{STk}$ and $\overline{kw}$. After receiving the search request $\overline{STk}$, the server then computes a predefined term, $e(\overline{STk}, \widetilde{STk})$, and if the result matches $\overline{kw}$, then the server sends the user the matched data file. The working principle is as the following. If a data file has the matched keyword, then

$$e(\overline{STk}, \widetilde{STk}) = e\left(H(w)^{sk}, g^{\frac{r}{sk}}\right) = e(H(w), g)^{sk*\frac{r}{sk}} = e(H(w), g)^r = \overline{kw}.$$

Since $\overline{STk}$, $\widetilde{STk}$, and $\overline{kw}$ are encrypted by the owner (using $sk$ or $r$), the server does not know who or what keyword is involved, but still can locate the matched data from the encrypted database. Note that the decryption processes in other layers in principle follow the same bilinear zero-knowledge principle. Details are elaborated later.

2.3 Complexity Assumptions

(1) Discrete logarithm problem assumption

Let G denote a cyclic multiplicative group with prime order $q$, and g be a generator of G. The discrete logarithm problem assumption is that given $(g, g^u)$, no polynomial-time algorithm can compute the value "$u$" with more than a negligible advantage.

(2) Decisional Bilinear Diffie-Hellman problem assumption [16]

Let G and GT denote two cyclic multiplicative groups with prime order $q$, g be a generator of G, and e be a bilinear mapping function: G × G → GT. Let $a, b, c \in Z_q$ be randomly chosen to produce values $(g, g^a, g^b, g^c)$. Note that one can easily compute $e(g,g)^{ab}$ as it is equal to $e(g^a, g^b)$.

### III. System Model

In this section, we introduce the basic system architecture about our proposed system scheme. In Fig. 2, we list all participants involved in the system and the relationship in the data sharing process.

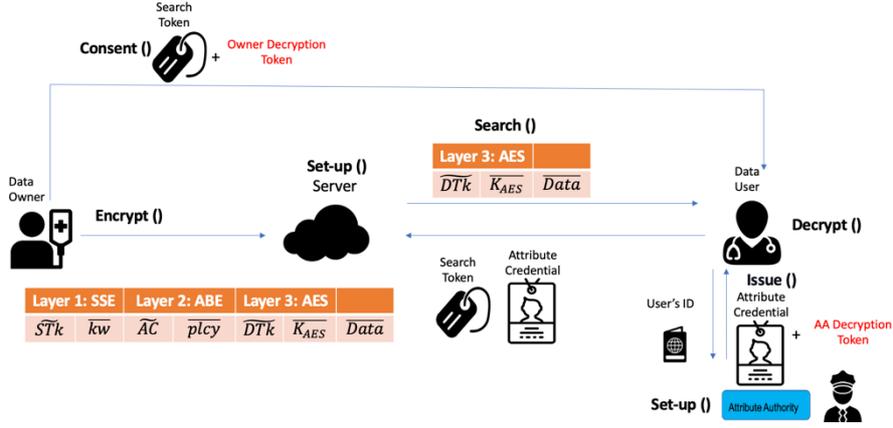

Fig. 2. The proposed 3LSAA data transaction system flow.

The proposed 3LSAA system involves four participants as shown in Fig. 2 and the flow is explained in the following. First, the server is an agent to set up public parameters using a **set-up** function, and the Attribute Authority (AA) sets up attribute key pairs using its own **set-up** function. Second, the owner sends the AES-encrypted data, $\overline{Data}$, along with bilinear cryptography elements to the server, using an **Encrypt** function. Third, the user obtains the search token, $\overline{STk}$, from the owner through the **Consent** function and also obtains the attribute credentials (AC) with a nonce from the Attribute Authority through the **Issue** function, and then sends to the server for data request. Fourth, the server processes the request from a user, through the **Search** function, and return to the user the matched encrypted data along with decryption elements. After receiving the matched results from the server, the user recovers the AES key using the **Decrypt** function, based on the owner issued decryption token $DTk_o$ and the AA issued decryption token $DTk_{AA}$.

Besides data transaction process, the data owner re-encryption (update) process is shown in Fig. 3.

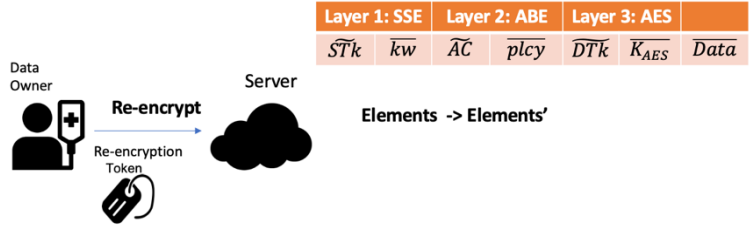

Fig. 3. The proposed re-encryption (update) model.

The purpose of the owner re-encryption process is to update the original bilinear cryptography elements, the settings of keyword and policy by issuing an owner generated Re-encryption token RTk. The re-encrypt process has 3 steps, verification, keyword update, and attribute update. The verification step is to check the qualification of RTk and prevent illegal update from attackers.

3.1 PARTICIPANTS

Here, we define each participant in the system:

**Data owner.** The data owner is a data producer, willing to share (trade) his data to others. For data preparation, the data owner first performs AES encryption of the data content, prepare keywords, ownership information (SSE) along with access policy structure (ABE), and then upload the encrypted data to the cloud server. Upon a qualified user's data request, the owner sends the user a search token with a specific data range and the decryption token for local decryption. The data owner is the only one who can update the associated data bilinear cryptography elements in the server. When needed, the owner simply sends to the server an authorized re-encryption token and have the server update data-tagged encrypted keywords, access policy, and the decryption element.

**Data user.** A data user is one who intends to access and use an owner's data for a purpose, such as research or targeted advertising. For the data request process, the data user asks for the search token and owner decryption token for targeted data from the data owner, and at the same time apply for attribute credentials and Attribute Authority (AA) decryption token from the attribute authorities. Then, after the data user sending to the cloud server the search tokens and attribute credentials, the server returns the matched result. Finally, the data user locally recovers the AES key using the owner issued decryption token and AA issued decryption token to decrypt the encrypted data.

**Cloud server.** The cloud server serves as an exchange agent between data owners and users. The server is responsible for keeping the encrypted data and processing bilinear cryptography elements updates for owners, executing request verification and data matching and returning matched data for and to qualified users. In this paper, we assume that the cloud server is honest-but-curious, and will honestly perform data update or access requests, but may try to poke and peek data stored and extract information from usage records.

**Attribute authority (AA).** An attribute authority serves as a public agency to authenticate credentials for users and issues certified credentials to qualified users. Typical AAs include government organizations, universities, or companies, et cetera. Only the users whose certified credentials match the owner delegated access policy on server receive matched data and can decrypt received data locally.

3.2 SYSTEM SCHEME, DEFINITIONS

In this section, we explain the system functions and terms used in Fig. 1 and Fig. 2. Note that for brevity, in the following we use the convention, *Executing Participant: Function (Inputs) → Receiving Participant: (Outputs)*, to indicate that the *Executing Participant* executes the *Function* by taking the *Inputs* parameters and then delivers the *Outputs* to the *Receiving Participant*. For clarity, for parameters, such as the decryption tokens from owner or AA, we use same parameter name to denote that both serve similar purpose, but apply different subscript to indicate the difference of source. For instance, for subscript we use *o* for owner, *u* for user, *AA* for attribute authority, and *S* for server. Therefore, $DTk_o$ means that the decryption token is from the owner.

The proposed 3-layer SSE-ABE-AES (3LSAA) system, as the name implies, has three working layers and each layer requires two cryptography elements installed in the server for each piece of data as shown in Fig. 4. We have previously explained the algorithm of the first layer, the SSE layer for keyword search. The second and third layers, ABE and AES, practically follow the same working principle.

| Layer 1: SSE | | Layer 2: ABE | | Layer 3: AES | | |
|---|---|---|---|---|---|---|
| $\widetilde{STk}$ | $\overline{kw}$ | $\widetilde{AC}$ | $\overline{plcy}$ | $\widetilde{DTk}$ | $\overline{K_{AES}}$ | $Data$ |

Fig. 4. The bilinear cryptography elements in the proposed 3LSAA model.

As for the bilinear cryptography elements from the owner for each piece of data, the SSE layer has search token transferor ($\widetilde{STk}$) and tagged-keyword ($\overline{kw}$) pair, the ABE layer has attribute credential transferor ($\widetilde{AC}$) and encrypted policy ($\overline{plcy}$) pair, and finally the AES key recovery layer has decryption token transferor ($\widetilde{DTk}$) and encrypted AES key ($\overline{K_{AES}}$) pair. After recovering the AES key, the user can decrypt the ciphertext and recover the original data. Since each encryption layer is independent of others, we discuss the operation of each layer separately. First, after introducing the SSE parameters, we discuss how the SSE's **Encrypt** of owner and **Search** function of server work and prove the correctness of the computation process. Then, we continue on the next layer and the last layer. Note that the owner generates all bilinear cryptography elements at the same time while the server processes SSE and ABE concurrently.

Since the advanced 3LSAA functions are more complicated, we first explain the basic computation process of SSE-ABE-AES then discuss the advanced features in latter sections.

## 3.3 BASIC 3LSAA SYSTEM, ALGORITHMS, KEYS, SECRETS

AES key ($K_{AES}$): A unique AES symmetric encryption key is used by the owner to encrypt each piece of data and the AES-encrypted ciphertext $\overline{Data}$ of the data is shown in Fig. 4. Specifically, we may use $K_{AES\_j}$ to represent the unique AES key for the owner's *j*-th data.

Owner Secret keys (*sk*): Each owner prepares a secret key *sk* for signing the search token.

**Server: Setup ($1^\lambda$) → Server: (*Param*).** The setup function of the cloud server is to provide the necessary **public parameters** (*Param*), by taking a computational complexity parameter $\lambda$. The unary representation $1^\lambda$ represents a string of $\lambda$ binary digits. For the security complexity parameter $\lambda$, an adversary requires $O(2^\lambda)$ effort to break the system. The cloud server also needs to announce to the public the cryptographic algorithms and the corresponding parameters used. The parameters include cyclic groups (G, GT), bilinear map function (*e*), generator of G (*g*), and hash functions (*H*). Let G and GT denote two cyclic groups with prime order *q*, the bilinear map function e: G × G → GT. The hash function H: $\{0,1\}^* \to$ G.

### 3.3.1 Basic SSE Computations

With the basic SSE computation explained in Fig. 1, we discuss the output of each function below.

**Owner: Encrypt (*sk*, $K_{AES}$, *Data*) → Server: ($\widetilde{STk}, \overline{kw}, \overline{Data}$).** The owner prepares the search token transferor ($\widetilde{STk}$) and the tagged-keyword ($\overline{kw}$) for each piece of *Data* using the owner secret key *sk*, a unique random AES key $K_{AES}$, based on the server public parameters (bilinear map function, *e*, generator of G, *g*, and hash functions, *H*). Basically, $\widetilde{STk} = g^{\frac{r}{sk}}$, $\overline{kw} = e(H(w), g)^r$, and $\overline{Data}$ is the AES encrypted *Data* using $K_{AES}$. On completion, the owner sends $\widetilde{STk}, \overline{kw},$ and $\overline{Data}$ to the server. Note that the owner may have the owner's ID as a keyword too, or $w = ID_o$, then the matching will enforce only the owner's data be located.

**Owner: Consent (*sk*, *w*) → User: ($\overline{STk}$).** An owner applies the consent function to produce a search token ($\overline{STk}$) with a keyword *w* using the owner secret key *sk*. Here, we have $\overline{STk} = H(w)^{sk}$. After completion, the owner sends the search token $\overline{STk}$ to the owner-consented user.

**Server: Search ($\widetilde{STk}, \overline{kw}, \overline{STk}$).** The server takes the search token ($\overline{STk}$) submitted from the data user, along with the pre-installed 3LSAA bilinear elements ($\widetilde{STk}, \overline{kw}$) and performs SSE data searching process. The computation follows our designed protocol and if the data file has the matched keyword, then,

$$e(\overline{STk}, \widetilde{STk}) = e\left(H(w)^{sk}, g^{\frac{r}{sk}}\right) = e(H(w), g)^{sk*\frac{r}{sk}} = e(H(w), g)^r = \overline{kw}.$$

Therefore, any file that satisfies the matching condition $e(\widetilde{STk}, \widetilde{STk}) = \overline{kw}$ will be located. Later in Section 3.4, we will discuss the advanced function that allows the owner to restrict only certain files in the search range.

3.3.2 Basic ABE Computations

The basic ABE process to be discussed below works similarly as the SSE process.

**Attribute Authority: Setup ($1^\lambda$) → Attribute Authority: (*ASK*, *APK*).** The attribute authority uses the setup function to prepare attribute key pairs, i.e., attribute secret key *ASK* and attribute public key *APK*, for attributes authentication management for each specific attribute, based on the security complexity parameter $\lambda$. The *ASK* is to authenticate user's attribute and sign an attribute credential to the user. The data owner uses the *APK*, the public key of the corresponding *ASK*, to set up the ABE bilinear encryption elements. In practice, the AA chooses a random number $a_i$ to be the *ASK*, and computes $\bar{a}_i = g^{1/a_i}$ to be the *APK* for the attribute $A_i$.

**Owner: Encrypt (*Attribute*, *APK*) → Server: ($\widetilde{AC}, \overline{plcy}$).** For the basic ABE process, the owner has to define the credential transferor, $\widetilde{AC}$, and access policy, $\overline{plcy}$, in the server for user access. The purpose of $\widetilde{AC}$ is similar to $\widetilde{STk}$ in SSE, and the purpose of $\overline{plcy}$ is similar to $\overline{kw}$. Here we use an example with two attributes, A1 and A2, to illustrate the operation. Assume that $\bar{a}_1$ and $\bar{a}_2$ are the attribute public keys and $s_1$ and $s_2$ are two owner-selected random numbers corresponding to A1 and A2. For the policy that is to qualify users with A1 AND A2, the owner sets

$$\widetilde{AC_{A1}} = (\bar{a}_1)^{s_1} (= g^{\frac{s_1}{a_1}})$$
$$\widetilde{AC_{A2}} = (\bar{a}_2)^{s_2} (= g^{\frac{s_2}{a_2}})$$

and

$$\overline{plcy} = e(g,g)^{s_1+s_2}.$$

Note that the AND relationship in the policy is equivalent to the plus relationship in the exponents. Then the owner prepared $\widetilde{AC}$ and $\overline{plcy}$ are sent to the server for later access control.

**Attribute Authority: Issue (*ASK*) → User: ($\overline{AC}$).** The attribute authenticator AA verifies the requesting user's attribute and uses AA's attribute secrete key *ASK* to sign an attribute credentials $\overline{AC}$, or $\overline{AC} = g^{ASK}$, to the qualified user. Note that each attribute requires a specific master secrete key. Therefore, if there are two attributes, A1 and A2, managed by AA, and the corresponding attribute secrete keys are $a_1$ and $a_2$, then $\overline{AC_{A1}} = g^{a_1}, \overline{AC_{A2}} = g^{a_2}$.

**Server: Search ($\widetilde{AC}, \overline{plcy}, \overline{AC}$).** The server takes the attribute credentials ($\overline{AC}$) submitted by the user and checks against with the owner installed bilinear elements ($\widetilde{AC}, \overline{plcy}$) for ABE fine-grained access control. We use the same example with two attributes, A1 and A2, to illustrate the operation. The server first compute

$$e\left(\overline{AC_{A1}}, \widetilde{AC_{A1}}\right) = e\left(g^{a_1}, g^{\frac{s_1}{a_1}}\right) = e(g,g)^{a_1 * \frac{s_1}{a_1}} = e(g,g)^{s_1}$$

$$e\left(\overline{AC_{A2}}, \widetilde{AC_{A2}}\right) = e\left(g^{a_2}, g^{\frac{s_2}{a_2}}\right) = e(g,g)^{a_2 * \frac{s_2}{a_2}} = e(g,g)^{s_2}$$

After finishing the attribute computations, the server merges the results based on the policy structure (A1 AND A2):

$$e(g,g)^{s_1} * e(g,g)^{s_2} = e(g,g)^{s_1+s_2}$$

or

$$e\left(\overline{AC_{A1}}, \widetilde{AC_{A1}}\right) * e\left(\overline{AC_{A2}}, \widetilde{AC_{A2}}\right) = \overline{plcy}.$$

The equality holds only if all ACs are valid. The above computation method works well for basic access control purpose but it cannot defend collusion. Someone may combine $\overline{AC}$'s from different sources to attack the ABE. The anti-collusion solution will be discussed in the advanced ABE part in the Section 3.4.

3.3.3 Basic AES Key Recovery

After the server finishes the SSE-ABE process in above, a valid user will receive $\overline{Data}$ along with decryption transferor $(\overline{DTk})$, and encrypted AES key $(\overline{K_{AES}})$, which is able to recover the corresponding AES for $\overline{Data}$ decryption. Similar to the SSE-ABE process, a valid user is issued a decryption token $\overline{DTk_o}$ from the owner and another decryption token $\overline{DTk_{AA}}$ from AA, which are computed with the transferor $(\overline{DTk})$. For the decryption token $DTk_o$, the subscript $o$ indicates that it is from the owner and for $DTk_{AA}$ the subscript $AA$ indicates the source is from AA. Note that both the owner and AA each applies a different secrete key from that used in SSE-ABE for signing $\overline{DTk_o}$, with purpose to avoid unauthorized AES key recovery using tokens collected from SSE-ABE. Therefore, we use the subscript $DTk$ in $SK_{DTk}$ and $ASK_{DTk}$ to indicate that they are for AES key recovery.

Owner Secret keys ($SK_{DTk}$): Each owner prepares another secret key $SK_{DTk}$, for generating decryption token $DTk_o$.

**Owner: Consent ($SK_{DTk}$) → User: ($\overline{DTk_o}$).** The owner signs $\overline{DTk_o} = g^{SK_{DTk}}$. Note that the Consent functions for $\overline{STk}$ and $\overline{DTk_o}$ are concurrently processed.

**Attribute Authority: Setup ($1^\lambda$) → Attribute Authority: ($ASK_{DTk}$, $APK_{DTk}$).** The attribute authority also installs a separate attribute key pair, $ASK_{DTk}$ and $APK_{DTk}$ for attributes authentication management in the key recovery process. Similar to the ABE, the AA chooses a random number $a'_i$ to be the $ASK_{DTk}$, and computes $\bar{a}'_i = g^{1/a'_i}$ to be the $APK_{DTk}$ for each attribute $A_i$.

**Attribute Authority: Issue** ($ASK_{DTk}$) → **User:** ($\overline{DTk_{AA}}$). The AA uses $ASK_{DTk}$ to sign $\overline{DTk_{AA}} = g^{ASK_{DTk}}$ corresponding to $\overline{AC} = g^{ASK}$. Each attribute requires a specific master secrete key. Therefore, if there are two attributes, A1 and A2, managed by AA, and the corresponding attribute secret keys are $a'_1$ and $a'_2$, then $\overline{DTK_{AA_{A1}}} = g^{a'_1}, \overline{DTK_{AA_{A2}}} = g^{a'_2}$.

**Owner: Encrypt** ($SK_{DTk}$, $APK_{DTk}$) → **Server:** ($\widetilde{DTk}$, $\overline{K_{AES}}$). We again use the access policy A1 AND A2 to illustrate the process. Similar to the ABE process, we set

$$\widetilde{AC_{A1}} = (\bar{a}_1)^{s_1} = g^{\frac{s_1}{a_1}}$$
$$\widetilde{AC_{A2}} = (\bar{a}_2)^{s_2} = g^{\frac{s_2}{a_2}}$$
$$\overline{plcy} = e(g,g)^{s_1+s_2}$$

The AND relationship in the policy is equivalent to plus relationship in the exponents. The AES key recovery process is to prove the user's qualification about SSE and ABE. In other words, the user shall be simultaneously eligible for both SSE and ABE. Besides the attribute A1 and A2, we can regard the SSE as the third attribute and the policy becomes (A1 AND A2 AND SSE). Let r be a random number used for SSE, the we now have $\overline{plcy} = e(g,g)^{s_1+s_2+r}$. Nevertheless, to avoid someone from utilizing the parameters in SSE and ABE, we shall use a new set of random number $s_1', s_2', r'$ to set a new $\overline{plcy} = e(g,g)^{s'_1+s'_2+r'}$ and use it to encrypt the AES key ($K_{AES}$) as

$$\overline{K_{AES}} = K_{AES} * e(g,g)^{s_1'+s_2'+r'}.$$

If a user is valid for both SSE and ABE, she can compute the value $e(g,g)^{s_1'+s_2'+r'}$ and use it to divide $\overline{K_{AES}}$ and then recovers AES key ($K_{AES}$), i.e., $K_{AES} = \overline{K_{AES}}/(e(g,g)^{s_1'+s_2'+r'})$.

In SSE and ABE, the owner prepares

$$\widetilde{STk} = g^{\frac{r}{sk}} \text{ and } \widetilde{AC} = (\bar{a}_i)^{s_i} = g^{\frac{s_i}{a_i}}.$$

Similarly, the owner now prepares $\widetilde{DTk} = g^{\frac{r'}{sk'}}$ for $\overline{DTk_o}$, and $\widetilde{DTk_{A_i}} = (\bar{a}_i')^{s_{i'}} = g^{\frac{s_{1'}}{a_{1'}}}$ for $\overline{DTk_{AA_{A_i}}}$

**User: Decrypt** ($\overline{DTk_o}$, $\overline{DTk_{AA}}$, $\widetilde{DTk}$, $\overline{K_{AES}}$) → **User:** ($K_{AES}$). After receiving the matched data from the cloud server, the user can recover the AES key ($K_{AES}$) for decrypting the received data by combining the $\overline{DTk_o}$, $\overline{DTk_{AA}}$ and $\widetilde{DTk}$.

To locally prove the validity of SSE using $\overline{DTk_o}$:

$$e(\overline{DTk_o}, \widetilde{DTk}) = e\left(g^{sk'}, g^{\frac{r'}{sk'}}\right) = e(g,g)^{sk'*\frac{r'}{sk'}} = e(g,g)^{r'}$$

To locally prove the validity of ABE using $\overline{DTk_{AA}}$:

$$e(\overline{DTk_{AA_{A1}}}, \widetilde{DTk_{A1}}) = e\left(g^{a1'}, g^{\frac{s1'}{a1'}}\right) = e(g,g)^{a1'*\frac{s1'}{a1'}} = e(g,g)^{s1'}$$

$$e(\overline{DTk_{AA_{A2}}}, \widetilde{DTk_{A2}}) = e\left(g^{a2'}, g^{\frac{s2'}{a2'}}\right) = e(g,g)^{a2'*\frac{s2'}{a2'}} = e(g,g)^{s2'}$$

After validating separately SSE and ABE, the user computes "SSE AND ABE":

$$e(\overline{DTk_o}, \widetilde{DTk}) * e(\overline{DTk_{AA_{A1}}}, \widetilde{DTk_{A1}}) * e(\overline{DTk_{AA_{A2}}}, \widetilde{DTk_{A2}})$$

or

$$e(g,g)^{r\prime} * e(g,g)^{s1\prime} * e(g,g)^{s2\prime} = e(g,g)^{s1\prime+s2\prime+r\prime}$$

With the result, the user computes $\overline{K_{AES}}/e(g,g)^{s\prime_1+s\prime_2+r\prime} = K_{AES}$ and recovers AES key ($K_{AES}$). The AES key is then used to decrypt $\overline{Data}$.

### 3.4 ADVANCED 3LSAA SYSTEM, ALGORITHMS, KEYS, SECRETS

For the advanced 3LSAA, the system flow is the same as the basic flow but we incorporate some advanced functions to solve the privacy and other issues mentioned previously. We first review the issues and then present the solutions.

#### 3.4.1 Advanced SSE Computations

First, we review the basic SSE process:

**Server: Setup ($1^\lambda$) → Server: (*Param (e, g, H)*).**

**Owner: Consent (*sk, Keyword (w)*) → User: ($\widetilde{STk} = H(w)^{sk}$).**

**Owner: Encrypt (*sk, Nonce (r)*) → Server: ($\widetilde{STk} = g^{\frac{r}{sk}}, \overline{kw} = e(H(w),g)^r$).**

**Server: Search ($\widetilde{STk}, \overline{kw}, \widetilde{STk}$).** Check if $e(\overline{STk}, \widetilde{STk}) = \overline{kw}$.

The server uses the token $\overline{STk}$ to locate all data indiscriminately that contain the supplied keyword $w$. For the advanced feature, we allow the owner to limit the user to search only a partial set of the data. To realize the partial set search, all owners in the system shall first classify their data into $n$ sets, where $n$ is determined by the owner, and use $i \in \{1, 2, \ldots, n\}$ for data set index. Then for attribute authority *AA*, we have each data set $i$ specific $ASK_i$, $APK_i$ used in the ABE process, and the server also prepares a specific $PK_i$ for data set $i$. Then we have

**Server: Setup ($1^\lambda$) → Server: (*Param (*public keys*, PK)*).**

Let $PK_i = \{g_i\} = \{g^{(a_i)}\}$, where $a_i$ is a server chosen random number for data set $i$. For all owners, they use the same $PK_i$ to classify their own i-th data.

Now suppose that the owner consents the user to access a partial data set $S$, then instead of the basic $\overline{STk} = H(w)^{sk}$, now the owner computes the advanced subset search token $\overline{STk}'$ as

$$\overline{STk}' = \prod_{i \in S} g_i^{sk} * H(w)^{sk} = \prod_{i \in S} g_i^{sk} * \overline{STk},$$

where $sk$ is the owner's secret key and $\prod_{i \in S} g_i^{sk}$ is the subset search token modifier.

Then the new search computation is:

$$e(\overline{STk}', \widetilde{STk}) = e\left(\prod_{i \in S} g_i^{sk} * H(w)^{sk}, g^{\frac{r}{sk}}\right)$$

$$= e\left(\prod_{i \in S} g_i, g\right)^{sk*\frac{r}{sk}} * e(H(w),g)^{sk*\frac{r}{sk}} = e\left(\prod_{i \in S} g_i, g^r\right) * e(H(w),g)^r$$

$$= e\left(\prod_{i \in S} g_i, g^r\right) * \overline{kw} = \overline{kw}',$$

where $e(\prod_{i \in S} g_i, g^r)$ is the subset keyword modifier.

We now discuss a defective design. One may consider to set the search token $\overline{STk}'$ to be $\prod_{i\in S} g_i^{\frac{sk}{r}} * \overline{STk}$, then the SSE computation will become $e(\overline{STk}', \widetilde{STk}) = e(\prod_{i\in S} g_i, g)^{\frac{sk}{r}*\frac{r}{sk}} * e(H(w), g)^{sk*\frac{r}{sk}} = e(\prod_{i\in S} g_i, g) * \overline{kw}$, and $e(\prod_{i\in S} g_i, g)$ becomes the subset keyword modifier which can be conveniently computed with no additional parameters. However, for this case the issue is that the $\overline{STk}'$ needs to use same nonce (r) for all owner's data, and as a result, the server may know the ownership of data since all $\overline{kw}$ and $\widetilde{STk}$ are identical. Therefore, our proposed approach applies $\prod_{i\in S} g_i^{sk}$ as the search token modifier ($\widetilde{STk}$) and has an additional parameter keyword modifier ($\widehat{kw} = g^r$) prepared by the owner for server's computation. In advanced 3LSAA, we apply the same cryptographic computation on the same element in the basic approach, but with a modifier. For convenience, we adopt the circumflex on the variable name, e.g., $\widehat{kw}$, to indicate that it is a modifier and add an apostrophe to indicate modified elements, e.g. $\overline{STk}'$ or $\overline{kw}'$. We list in Fig. 5 the cryptography elements for the advanced SSE installed in the server along with $\overline{Data}$.

**Layer 1: SSE**

$$\widetilde{STk} = g^{r/sk} \quad \widehat{kw} = g^r \quad \overline{kw} = e(H(w), g)^r$$

Fig. 5. The bilinear cryptography elements for the advanced SSE.

In summary, to perform subset search, the owner first determines the data subset $S$ for search and based on the targeted keyword generates a modified search token $\overline{STk}'$. Then the user submits $\overline{STk}'$ and data subset $S$ to the server who then computes the modified $\overline{kw}'$ using subset keyword modifier $e(\prod_{i\in S} g_i, g^r)$. Finally, the server verifies if $e(\overline{STk}', \widetilde{STk}) = \overline{kw}'$ to identify matched data.
If a user lies and tells the server a different data set $S'$ instead of $S$, then the search shall fail since
$$e(\overline{STk}', \widetilde{STk}) = e(\prod_{i\in S} g_i, g^r) * \overline{kw} \neq e(\prod_{i\in S'} g_i, g^r) * \overline{kw}.$$

Consequently, a requesting user submits search token $\overline{STk}'$ and the search data set S for the server to perform the search function as the following,
**Server: Search ($\widetilde{STk}$, $\overline{kw}$, $\widehat{kw}$, $\overline{STk}'$, data set ($S$)).**
Check if $e(\overline{STk}', \widetilde{STk}) = e(\prod_{i\in S} g_i, \widehat{kw}) * \overline{kw} = \overline{kw}'$.

### 3.4.2 Advanced ABE Computations
In the basic ABE, AA chooses a random number $a_i$ to be the attribute secrete key (*ASK*), and generates attribute public key (*APK*) $\bar{a}_i = g^{a_i}$ for the attribute A$_i$. The specific computations are
**Attribute Authority: Setup (1$^\lambda$) → Attribute Authority: (*ASK ($a_i$)*, *APK ($g^{a_i}$)*).**
**Attribute Authority: Issue (*ASK ($a_i$)*) → User: ($\overline{AC} = g^{a_i}$).**
**Owner: Encrypt (*APK ($g^{a_i}$)*, *Nonce ($s_i$)*) → Server: ($\widetilde{AC} = g^{\frac{s_i}{a_i}}$, $\overline{plcy} = e(g,g)^{\Sigma_{i\in P} s_i}$).**
**Server: Search ($\widetilde{AC}$, $\overline{plcy}$, $\overline{AC}$).** Check the validity of the pairing of $\overline{AC}$ and $\widetilde{AC}$ based on the policy structure:

$$\prod_{i \in P} e(\overline{AC_{Ai}}, \widetilde{AC_{Ai}}) = \prod_{i \in P} e\left(g^{ai}, g^{\frac{si}{ai}}\right) = \prod_{i \in P} e(g,g)^{ai*\frac{si}{ai}}$$

$$= \prod_{i \in P} e(g,g)^{si} = e(g,g)^{\Sigma_{i \in P} s_i} = \overline{plcy}.$$

An issue of the basic ABE is that it cannot defend collusion. Someone may combine $\overline{AC}$'s from different sources to attack the ABE. Since the $\overline{AC}$ is not user specific, the server cannot distinguish the source of $\overline{AC}$. Our solution is to have AA issue $\overline{AC}$ based on user's real identity or $GID$ and have the user select a nonce $r$ in each request to produce different $\overline{AC}$ to be anonymous. Given the attribute A$_i$ and user's GID and nonce $r$, the AA computes

$$\overline{AC'} = g^{ai} * H(GID)^{r*ai} = \overline{AC} * H(GID)^{r*ai}.$$

Now with the credential modifier $H(GID)^{r*ai}$, the server cannot link the request to the user identity.

In contrast to the basic ABE $\overline{plcy} = e(g,g)^{\Sigma_{i \in P} s_i}$, in the advanced ABE the server computes

$$\prod_{i \in P} e(\overline{AC'_{Ai}}, \widetilde{AC_{Ai}}) = \prod_{i \in P} e(g^{ai} * H(GID)^{r*ai}, g^{\frac{si}{ai}}) = \prod_{i \in P} e(g * H(GID)^r, g)^{si}$$

$$= \prod_{i \in P} e(g,g)^{si} e(H(GID)^r, g)^{si} = e(g,g)^{\Sigma_{i \in P} s_i} \prod_{i \in P} e(H(GID)^r, g^{si})$$

$$= \overline{plcy} * \prod_{i \in P} e(H(GID)^r, g^{si}) = \overline{plcy'},$$

where $P$ is the set of valid attributes in the policy and $\prod_{i \in P} e(H(GID)^r, g^{si})$ is the policy modifier.

Therefore, for advanced ABE, the server needs to compute the modified policy $\overline{plcy}'$ using the policy modifier $\prod_{i \in P} e(H(GID)^r, g^{si})$ and verify the equality $\prod_{i \in P} e(\overline{AC'_{Ai}}, \widetilde{AC_{Ai}}) = \overline{plcy'}$. In practice the owner shall prepare $g^{si}$ as an additional modifier ($\widehat{plcy} = g^{si}$) at the setup time for the server. Then for a data request, the user shall submit the value $H(GID)^r$ along with the AA issued $\overline{AC'}$ to the server for verification. The server shall know the policy $P$ to make correct pairing between $\overline{AC_{Ai}}$ and $\widetilde{AC_{Ai}}$. The advanced ABE parameters are summarized in Fig. 6.

| Layer 2: ABE | | |
|---|---|---|
| $\widetilde{AC_{Ai}} = g^{si/ai}$ | $\widehat{plcy} = g^{si}$ | $\overline{plcy} = e(g,g)^{\Sigma_{i \in P} s_i}$ |

Fig. 6. The bilinear cryptography elements for the advanced ABE.

Finally, the server performs the access control function as the following,

**Server: Search ($\widetilde{AC}$, $\overline{plcy}$, $\overline{AC'}$, $\widehat{plcy}$, $H(GID)^r$).** The server checks if

$$\prod_{i \in P} e(\overline{AC'_{Ai}}, \widetilde{AC_{Ai}}) = \overline{plcy} * \prod_{i \in P} e(H(GID)^r, \widehat{plcy}) = \overline{plcy'},$$

where $P$ is the set of valid attributes in the policy.

3.4.3 Advanced AES Key Recovery

The advanced AES algorithm is based on the advanced SSE-ABE approach. First, the advanced subset decryption tokens from the owner is $\overline{DTk_o}' = \prod_{i \in S} g_i^{sk\prime} * g^{sk\prime} = \prod_{i \in S} g_i^{sk\prime} * \overline{DTk_o}$ and the advanced global identity decryption token from the attribute authority is $\overline{DTK'_{AA_{A_l}}} = g^{ai\prime} * H(GID)^{r'*ai\prime} = \overline{DTK_{AA_{A_l}}} * H(GID)^{r'*ai\prime}$.

The validity of SSE is verified based on $\overline{DTk'_o}$:

$$e\left(\overline{DTk'_o}, \widehat{DTk_o} = g^{\frac{r'}{sk\prime}}\right) = e\left(\prod_{i \in S} g_i, g^{r'}\right) * e(g,g)^{r'}$$

The validity of ABE is verified based on $\overline{DTk'_{AA}}$:

$$\prod_{i \in P} e\left(\overline{DTK'_{AA_{A_l}}}, \widetilde{DTk_{AA_{A_l}}} = g^{s'_i * a'_i}\right) = e(g,g)^{\Sigma_{i \in P} s_i} * \prod_{i \in P} e(H(GID)^{r'}, g^{s'i})$$

Sine $K_{AES} = \overline{K_{AES}}/e(g,g)^{\Sigma_{i \in P} s_i + r'}$, the user may derive the value $e(g,g)^{\Sigma_{i \in P} s_i + r'}$ by canceling out $e(\prod_{i \in S} g_i, g^{r'})$ and $\prod_{i \in P} e(H(GID)^r, g^{s'i})$ from the multiplication of the above two results using the two decryption token modifiers, $\widehat{DTk_o} = g^{r'}$ and $\widetilde{DTk_{AA_{A_l}}} = g^{s'i}$.

| Layer 3: AES | | |
|---|---|---|
| $\widehat{DTk_o} = g^{r'/sk\prime}$ | $\widehat{DTk_o} = g^{r'}$ | $\overline{K_{AES}} = K_{AES} * e(g,g)^{\Sigma_{i \in P} s_i + r'}$ |
| $\widehat{DTk_{AA_{Ai}}} = g^{si\prime/ai\prime}$ | $\widehat{DTk_{AA_{Ai}}} = g^{s'i}$ | |

Fig. 7. The bilinear cryptography elements in the advanced AES Key Recovery.

In other words, we may compute

$$e(g,g)^{r'} = e\left(\overline{DTk'_o}, \widehat{DTk_o} = g^{\frac{r'}{sk\prime}}\right) / e\left(\prod_{i \in S} g_i, \widehat{DTk_o} = g^{r'}\right)$$

$$e(g,g)^{\Sigma_{i \in P} s'_i} = \prod_{i \in P} e\left(\overline{DTK'_{AA_{A_l}}}, \widetilde{DTk_{AA_{A_l}}} = g^{s'_i * a'_i}\right) / \prod_{i \in P} e\left(H(GID)^r, \widetilde{DTk_{AA_{A_l}}} = g^{s'i}\right)$$

With the result, the user computes $\overline{K_{AES}}/e(g,g)^{\Sigma_{i \in P} s'_i + r'} = K_{AES}$ and recovers AES key $K_{AES}$ to decrypt $\overline{Data}$.

3.5 Re-encryption (update) computations

Re-encryption process is to update the settings of keywords and policy by issuing an owner generated re-encryption token (RTk). Note that with the extended integrity requirement, the cloud server can grant only to the data owner the right to update the data properties. Therefore, to defend illegal update from attackers, the server needs to follow a verification process. Specifically, the verification step follows the advanced SSE process used for finding a specific owner's data:

$$e(\overline{STk'}, \widetilde{STk}) = e\left(\prod_{i \in S} g_i^{sk} * H(w)^{sk}, g^{\frac{r}{sk}}\right) = e\left(\prod_{i \in S} g_i, g^r\right) * e(H(w),g)^r = \overline{kw'}$$

In advanced SSE, the owner may set the owner's ID as a keyword, $w = ID_o$. In order to avoid an attacker reuses the same $w = ID_o$ in both SSE and Re-encryption processes, we prepare another specific id as a keyword ($w = ID_{RTk}$), and generate a new encrypted keyword $\overline{kw_{RTk}} = e(H(w = ID_{RTk}), g)^r$ for requester's verification.

Then the verification process of *RTk* can be modified to:

$$e(\overline{RTk}, \widetilde{RTk}) = e\left(\prod_{i \in S} g_i^{sk} * H(w = ID_{RTk})^{sk}, g^{\frac{r}{sk}}\right) = e\left(\prod_{i \in S} g_i, g^r\right) * e(H(w = ID_{RTk}), g)^r$$

$$= e\left(\prod_{i \in S} g_i, g^r\right) * \overline{kw_{RTk}} = \overline{kw_{RTk}}'$$

Similar to the advanced SSE, we may confine the re-encryption operations to only a partial set of data S. Only after passing the verification, the server processes the re-encryption (update) request, for which the owner shall provide new bilinear cryptography elements for replacement.

**IV. System Analysis**

In this section, we analyze the security and time complexity of the proposed 3LSAA algorithm. We assume that the server is honest-but-curious, i.e., the server will correctly compute the pre-defined algorithm, but may try to learn the owner's or user's secret based from the usage traces. Besides the potential attacks from server, the user may try to access unauthorized data or even collude with other users. Based on these considerations, we prove the security and privacy protection of 3LSAA.

4.1.1   Advance SSE Complexity Analysis

For the SSE layer, the complexity of the owner **Encrypt** function is linear to the size of the data, since each piece of data needs a tuple of bilinear elements and the computation is constant.

For the owner **Consent** function, the complexity of the search token $\overline{STk}$ making depends on the size of data subset authorized, and is linear to the owner's maximum data size.

As for the server **Search** function, the server checks through all data in database, and hence the complexity is O(n), where n is the total amount of data in the server database. Note that, the search can be performed in parallel since the search computation for each piece of data is independent.

4.1.2   Advance SSE Security Analysis

With our approach, the user data accesses are limited to the owner approved scope. One important improvement for our proposed advanced SSE is that it allows the owner to limit the user to search only within a partial set of the data. Therefore, the user cannot access the data out of scope. Since the owner has imposed $\prod_{i \in S} g_i^{sk}$ with the owner's secret key $sk$ into the search token $(\overline{STk})$ in advance, the search scope is then confined to the data subset S. If the user declares a wrong set $S'$, the computation in the server side will fail to match any data.

4.2.1   Advance ABE Complexity Analysis

For the Attribute Authority **Issue** function, the computation is O(1), constant time, for each credential issue. If the server shall examine all attributes of a user, the computation complexity will be the number of attributes the server manages.

For the owner **Encrypt** function to define an access policy, the complexity is O(m) with m attributes for each piece of data, and the total computation complexity is O(m*n) for n pieces of data.

For the server **Search** function, the complexity of the server examination of policy depends on the number of attributes (m). In other words, the total computation complexity is O(m*n) for n pieces of data. Since the SSE's examination for each piece of data takes O(1) but the ABE takes O(m) to filter the database, Therefore SSE-ABE is a more efficient computation order for server search computation.

4.2.2   Advance ABE Security Analysis

Our approach allows no users collusion of Attribute Credential for access policy attacks.
If a user uses a credential with different *GID* from others, then the user shall submit the value $H(GID)^r$ along with the AA issued $\overline{AC'}$ to the server for verification, which is created using a different *GID,* hence the verification will fail. Additionally, we apply a random nonce (r) in making the $\overline{AC_{At}}$ and prevent the server from knowing the user's identity for privacy protection.

4.3 Advanced AES Key Recovery and Re-encryption (update)

Since the AES key recovery leverages the SSE-ABE process and Re-encryption (update) utilizes the advanced SSE approach, the computation complexity and security protection essentially are the same as that of the SSE-ABE and the advanced SSE methods. However, to prevent the server from reusing the elements in SSE-ABE process to attack the advanced AES key recovery and re-encryption, the owner applies a different secrete key and nonce for token encryption in the recovery and re-encryption and avoid the reuse attack.

**V.   Conclusion**

In this paper, we have proposed a new protocol, 3-layer SSE-ABE-AES (3LSAA) and effectively resolve the critical virtual ownership issues exists in the traditional centralized data-sharing server scheme. Our approach lets the data owners regain the actual access control and full data privacy protection. Essentially, we apply the zero-knowledge technique in each layer of the 3LSAA protocol and achieve extreme high security and privacy protection. Additionally, the approach realizes automatic access control management with convenient file search feature and great usability improvement. We also proposed an extended CIA model to evaluate data privacy and security level of the cross-organization system under an untrusted environment. The system complexity is linear to the amount of data managed and hence is expected to be practical for daily applications.